\documentclass[a4,aps,prd,12pt]{revtex4}

\usepackage{amsmath,amssymb}

\usepackage{color}
\usepackage[dvips]{graphicx}

\newcommand{\lsim}{\mathrel{\mathop{\kern 0pt \rlap
  {\raise.2ex\hbox{$<$}}}
  \lower.9ex\hbox{\kern-.190em $\sim$}}}
\newcommand{\gsim}{\mathrel{\mathop{\kern 0pt \rlap
  {\raise.2ex\hbox{$>$}}}
  \lower.9ex\hbox{\kern-.190em $\sim$}}}

\newcommand{\be}{\begin{equation}}
\newcommand{\ee}{\end{equation}}
\newcommand{\beqarr}{\begin{eqnarray}}
\newcommand{\eeqarr}{\end{eqnarray}}

\begin{document}


\title{Dark matter in the Kim-Nilles mechanism}



\author{Eung Jin Chun}
\affiliation{Korea Institute for Advanced Study\\
Hoegiro 87, Dongdaemun-gu, Seoul 130-722, Korea\\
Email) {ejchun@kias.re.kr}
 }

%

\begin{abstract}
The Kim-Nilles mechanism relates the $\mu$ term with the axion
scale $f_a$, leading to the axino-Higgsino-Higgs Yukawa coupling
of order $\mu/f_a$.  This can bring a dangerous thermal production
of axinos.  If the axino is stable, its mass has to be as small as
${\cal O}$(0.1keV), or the reheat temperature should be lower than
${\cal O}$(10GeV) taking the lower axion scale $10^{10}$GeV
in order not to overclose the Universe. If
the axino decays to a neutralino, the overproduced neutralinos can
re-annihilate appropriately to saturate the observed dark matter
density if the annihilation rate is of order
$10^{-8}\mbox{GeV}^{-2}$ for the axion scale larger than about $10^{11}$GeV.
Thus, a light Higgsino-like lightest
supersymmetric particle with a sizable bino mixture becomes a good
dark matter candidate whose nucleonic cross-section is larger than about
$10^{-45}$cm$^2$.
\end{abstract}


\maketitle


\section{Introduction}

The strong CP problem is elegantly solved by the axion solution leading to the dynamical
$\theta$ term:
\begin{equation} \label{aGG}
 {\cal L}^{\rm QCD}_a = {g_s^2 \over 32\pi^2} {a\over f_a} G^a_{\mu\nu} \tilde{G}_a^{\mu\nu},
\end{equation}
where $f_a$ is the axion decay constant allowed in the range of
$10^{10} \mbox{GeV} \lsim f_a \lsim 10^{12} \mbox{GeV}$
\cite{Kim08}. In the supersymmetric extension of the Standard
Model, the origin of  the supersymmetric Higgs mass term, $\mu$,
can be related to the axion scale through the realization of the
DFSZ axion \cite{dfsz}.  This is the Kim-Nilles mechanism
\cite{Kim84} which introduces the $\mu$ term interaction,
\begin{equation} \label{aHH}
 W_{KN} = \lambda {S^2 \over M_P} H_1 H_2,
\end{equation}
where $M_P=2.4\times10^{18}$ GeV is the reduced Planck mass.
Recall that $S$ contains the axion degree of freedom and generates
the axion scale through its vacuum expectation value at an
intermediate scale, $\langle S\rangle \sim f_a$, and thereby
induces  the $\mu$ term: $\mu \sim f_a^2/M_P$.   The superpartner
of the axion, the axino $\tilde{a}$, gets massive upon
supersymmetry breaking, and can be  the lightest supersymmetric
particle (LSP) to become a dark matter candidate.  However,
axinos can be efficiently produced from the thermal bath and
typically  their abundance is problematically large.

In the case of the KSVZ axion \cite{ksvz}, the axino-gluino-gluon interaction,
a supersymmetric counterpart of the axion-gluon-gluon interaction (\ref{aGG}),
\begin{equation} \label{taGG}
 {\cal L}^{QCD}_{\tilde{a}} = {g_s^2\over 32\pi^2}
 {1\over f_a}\, \tilde{a}\, \sigma^{\mu\nu} \tilde g^a
 G^a_{\mu\nu} + h.c. \,,
\end{equation}
is the main source for the thermal production of axinos either
through  freeze-out \cite{Raja91} or  through regeneration
\cite{Covi01,Brand04,Strumia10}. Such processes are so efficient
that the axino must be very light in order not to overclose the
Universe unless the reheat temperature is below the electroweak
scale to suppress its thermal production. Such a light axino
requires  special arrangements in the superpotential or
supersymmetry breaking parameters to suppress tree and loop
contributions to its mass which is generically of order  the
gravitino mass in gravity mediated supersymmetry breaking models
\cite{Goto92,Chun95}. On the other hand, smaller axino mass may
arise more naturally in gauge mediation models \cite{Chun99,Jeong11}.

\medskip

So far, the cosmic axino production has been studied extensively
concentrating on the KSVZ axion model. In this paper, we consider
the supersymmetric DFSZ axion model realized in the Kim-Nilles
mechanism extending limited discussions in
Refs.~\cite{Chun00,Choi11}. The DFSZ axino production is led by
the effective $\mu$ term interaction (\ref{aHH}) which, after the
field $S$ gets a vacuum expectation value,  induces the
axino-Higgsino-Higgs Yukawa term:
\begin{equation} \label{taHH}
  {\cal L}^{KN}_{\tilde a} = c_H{\mu\over f_a}\, \tilde{a}\,
  [ \tilde{H}_1 H_2 + \tilde{H}_2 H_1] + h.c.,
\end{equation}
where $c_H$ is a model-dependent parameter of order one determined
by the precise relation between $\langle S\rangle$ and $f_a$ and
the axino fraction in the fermion component $\tilde{S}$.

As will be discussed in detail, the DFSZ axino interaction
(\ref{taHH})  is stronger than the QCD interaction  (\ref{taGG})
at low reheat temperature and thus the DFSZ axinos are also
produced efficiently leading to a stringent mass bound or an
strong upper limit on the reheat temperature below the electroweak
scale when the axino is stable. On the other hand, if the DFSZ
axino decays to a Higgsino and a Higgs boson,
the decay occurs before thermal freeze-out of the
usual neutralino LSP dark matter for lower values of $f_a$. For
higher values of $f_a$, the DFSZ axino decays after the neutralino
freeze-out and decay-produced neutralinos may overclose the
Universe unless they re-annihilate sufficiently \cite{Choi08}. For
a typical choice of the parameters, the neutralino annihilation
rate of order $10^{-8}$ GeV$^{-2}$ turns out to lead to a right
amount of dark matter abundance, which makes a light Higgsino LSP
a natural dark matter candidate \cite{Baer11}. When the
Higgsino-like dark matter contains a sizable fraction of bino, its
spin-independent nucleonic scattering cross-section in the
decoupling limit of heavy Higgs bosons and sfermions is shown to
be of order $10^{-45}$cm$^2$  which is
within the reach of future direct detection experiments.

\section{Axino dark matter}

{\bf A light DFSZ axino:} Below the axion scale $f_a$,  the
axino-Higgsino-Higgs Yukawa interaction (\ref{taHH}) is generated
and plays a major role in producing the axino number density from
thermal regeneration. The axino number density  in unit of the
entropy density, $Y_{\tilde a}\equiv n_{\tilde a}/s$, is
determined by solving the Boltzmann equation:
\begin{equation} \label{BE}
 z s H {d Y_{\tilde a} \over d z } = \gamma\,,
\end{equation}
where $s=(2\pi^2/45)g_* T^3$,  $H=0.33\sqrt{g_*}\, T^2/M_P$  is
the Hubble parameter, $g_*$ is the number of the degrees of
freedom in thermal equilibrium, and  $z\equiv m/T$ with the
relevant mass parameter $m$ of the process $\gamma$. For our
calculation of the axino production, we will consider decay (or
inverse decay) processes coming from the $\mu$ term interaction
(\ref{taHH}) as their effect is expected to be larger than
scattering processes when decay channels are open \cite{Hall09}.

Let us first consider the decay process $H_1 \to \tilde a
\tilde{H}_1$  taking the Higgs mass $m_{H_1}$ larger than the
Higgsino mass $\mu$ before the electroweak symmetry breaking.
Integrating Eq.~(\ref{BE}), one finds
\begin{equation} \label{YaH}
 Y_{\tilde a} \approx {135 g \over 4 \pi^4 g_*^{3/2}} {\Gamma_{H_1} M_P \over m_{H_1}^2}
 \int_0^\infty z^3 K_1(z) dz \,,
\end{equation}
where  $g=4$ including the Higgs degrees of freedom and the final
axino number,   $\Gamma_{H_1} \approx c_H^2 (\mu/ f_a)^2
m_{H_1}/8\pi$ and $z=m_{H_1}/T$. This gives
\begin{equation}  \label{OmegaH}
 \Omega_{\tilde a}^{H} h^2 \approx 0.11
 \left(m_{\tilde a} \over 72 \mbox{keV} \right)
 \left( \mu \over 500 \mbox{GeV}\right) \left( 1 \mbox{TeV} \over m_{H_1}\right)
 \left( 10^{11} \mbox{GeV} \over f_a/c_H\right)^2 \,,
\end{equation}
for $g_*=200$.

When the temperature of the Universe falls below the critical
temperature,  $T_c\sim 100$ GeV, of the electroweak phase
transition, new decay channels can open. In the decoupling limit
of heavy Higgs bosons,  the coupling of the lightest Higgs boson,
$h$, resulting from the interaction (\ref{taHH}) is
\begin{equation} \label{LahH}
 {\cal L}_{\tilde a}^h = c_H {\mu\over f_a}  \, \tilde a [c_\beta \tilde H^0_1
 + s_\beta \tilde H^0_2] h + h.c.,
\end{equation}
where the angle $\beta$ is defined by $t_\beta = \langle H^0_2\rangle /\langle H^0_1\rangle$. Assuming almost degenerate Higgsinos (with the mass $m_{\tilde H} \approx \mu$)
decaying to $\tilde a +h$,  one obtains the axino abundance:
 \begin{equation} \label{YaHt}
 Y_{\tilde a} \approx {135 g \over 4 \pi^4 g_*^{3/2}} {\Gamma_{\tilde H} M_P \over m_{\tilde H}^2} \int_{z_c}^\infty z^3 K_1(z) dz \,,
\end{equation}
where $z_c=m_{\tilde H}/T_c$, $g=2$ and $\Gamma_{\tilde H}\approx c_H^2 (\mu/f_a)^2 m_{\tilde H}/16\pi$. Taking $m_{\tilde H} = \mu =500$ GeV and   $T_c=100$ GeV the integration in Eq.~(\ref{YaHt}) from $z_c=5$ gives
\begin{equation} \label{OmegaHt}
 \Omega_{\tilde a}^{\tilde H} h^2 \approx 0.11
 \left(m_{\tilde a} \over 300 \mbox{keV} \right)
 \left( \mu \over  m_{\tilde H}\right)^2
 \left( 10^{11} \mbox{GeV} \over f_a/c_H\right)^2 \,,
\end{equation}
for $g_*=100$. Another potentially important process for the DFSZ axino production comes from the supersymmetric top quark Yukawa coupling:
\begin{equation} \label{Latt}
 {\cal L}^t_{\tilde a} = c_t {m_t\over f_a}\, [\,\tilde a t \tilde t^c + \tilde a t^c \tilde t \,] +h.c.\,,
\end{equation}
where $c_t$ quantifies the axino fraction in the Higgsino $\tilde H_2$.  This opens up a new channel of the stop decay to an axino and the resulting axino population can be calculated  as in Eq.~(\ref{YaHt}) with a replacement of $\mu \to m_t$, $m_{\tilde H} \to m_{\tilde t, \tilde t^c}$ and $z_c=m_{\tilde t, \tilde t^c}/T_c$. One would get a similar result as in Eq.~(\ref{OmegaHt}) except a more suppression of $(m_t/m_{\tilde t, \tilde t^c})^2$ for $m_{\tilde t, \tilde t^c} \approx m_{\tilde H}$.  We will use the stop decay for the case of a heavy axino.

\medskip

The above results can be compared with the axino production from
the QCD interaction (\ref{taGG}) \cite{Strumia10}:
\begin{equation} \label{OmegaQCD}
 \Omega_{\tilde a}^{QCD}h^2 \approx 0.11
 \left(m_{\tilde a} \over 44 \mbox{MeV} \right)
 \left( T_R \over 10^3 \mbox{GeV}\right)
 \left( 10^{11} \mbox{GeV} \over f_a\right)^2 \,,
\end{equation}
where $T_R$ is the reheat temperature of the Universe. This shows that the KSVZ axino production is more effective than the DFSZ axino production for higher reheat temperature: $T_R> 10^{5\mbox{--}6}$ GeV for which the small loop factor ($g_s^2/32\pi^2$) of the QCD interaction term (\ref{aGG}) can be overcome.  In any case, the axino mass must be much smaller than the electroweak scale which requires some special arrangement in the supersymmetric axion sector as mentioned in the Introduction.

\medskip

{\bf A heavy DFSZ axino:} When the axino mass takes its typical value of the soft supersymmetry breaking scale in gravity mediation, the above strong bounds can be evaded if the axino population is diluted by entropy dumping or the axino decays to a lighter dark matter candidate like the gravitino \cite{Choi11,Cheung11}. The first possibility has been discussed in the context of thermal inflation \cite{Chun00}.  In this case, the axino number density is suppressed by the Boltzmann factor and thus is quite sensitive to the reheat temperature after (thermal) inflation or any entropy dumping.
To derive an appropriate reheat temperature for which the axino abundance saturates the required dark matter density of the Universe: $\Omega_{DM} h^2\approx 0.11$, let us take the axino production through the stop decay (\ref{Latt}) after the electroweak symmetry breaking. Modifying the result (\ref{YaHt}), the axino number density is given by
\begin{equation} \label{Yatt}
 Y_{\tilde a} \approx {135 g \over 4 \pi^4 g_*^{3/2}} {\Gamma_{\tilde t} M_P \over m_{\tilde t}^2} \int_{z_R}^\infty z^3 K_1(z) dz \,,
\end{equation}
where $g=2$, $\Gamma_{\tilde t}\approx c_t^2 (m_t/f_a)^2 m_{\tilde t}/8\pi$ and $z_R=m_{\tilde t}/T_R$. Here the integration is essentially determined by the value of $z_R$ and  the integration to the infinity is justified if the stop freeze-out temperature is smaller than $T_R$.  For $z_R\gg1$, the integration is well approximated  by $\sqrt{\pi/2} z_R^{5/2} \exp(-z_R)$.
Thus, the axino relic density produced from the stop decay at low reheat temperature becomes
\begin{equation} \label{Omegat}
 \Omega_{\tilde a}^{\tilde t} h^2 \approx 0.1 \left( m_{\tilde a} \over 500 \mbox{GeV}\right)
 \left( 1\mbox{TeV} \over m_{\tilde t}\right)
 \left( 10^{11} \mbox{GeV} \over f_a \right)^2 \left( z_R\over 20.2\right)^{5/2} e^{-z_R+20.2}
\end{equation}
for $m_t=172$ GeV, $g_*=100$ and $c_t=1$. Taking $f_a=10^{10}$ GeV, one gets $z_R=25.4$ implying that the dangerous heavy axino relic density can be suppressed sufficiently for $T_R \lsim 40$ GeV.
A stronger bound can be obtained, e.g., from the coupling (\ref{LahH}) if a heavy Higgsino decay channel to the axino is open.


\section{Light Higgsino dark matter}

A more interesting possibility appears when a heavy axino is
allowed to decay to the LSP without resorting to the above option
of low reheat temperature. The DFSZ axino can decay to a usual
neutralino through the $\mu$ term coupling (\ref{LahH}) leading to
overabundant population of the LSP dark matter. If the decay
occurs before freeze-out of the dark matter, such a overproduction
is harmless as the produced dark matter equilibrates and its relic
density is determined by the usual freeze-out calculation
\cite{Jungman95}. Although the axino decays later, the
overproduced dark matter population can get depleted appropriately
through a strong re-annihilation and thus can settle down to the
required value of $\Omega_{DM}h^2\approx 0.11$ \cite{Choi08}.

In fact, the Kim-Nilles mechanism provides a natural framework to realize such a re-annihilating dark matter scenario.
As we will see, the annihilation rate of a light Higgsino LSP of order
$10^{-8} \mbox{GeV}^{-2}$ and the DFSZ axino decay rate  turn out to be
in the right range to make a good dark matter candidate the Higgsino LSP produced
from the DFSZ axino decay $\tilde a \to h \tilde H $ for the axion scale of $f_a\approx 10^{11\sim12}$GeV.

\medskip

Let us first consider if the heavy DFSZ axino population can be large enough.
From the discussions in the previous section, the axino can be produced from the stop decay as in (\ref{Yatt}) but with the replacement of $z_R\to z_c \approx m_{\tilde t}/T_c$.
This gives $Y_{\tilde a} \approx 5\times 10^{-9}$ for $g_*=100$, $f_a=10^{11}$GeV,
$m_{\tilde t}=1$TeV and $T_c=100$ GeV. A more effective and inevitable process is the inverse decay
$h \tilde H \to \tilde a$.  From the couplings in Eq.~(\ref{LahH}), we find
\begin{eqnarray} \label{YahH}
 Y_{\tilde a} &\approx& {135  \over 4 \pi^4 g_*^{3/2}} {\Gamma_{\tilde a} M_P \over m_{\tilde a}^2} \int_{z_c}^\infty z^3 K_1(z) dz \nonumber\\
 &\sim& 10^{-7}\left(\mu \over 200 \mbox{GeV}\right)^2
 \left( 10^{11} \mbox{GeV} \over f_a/c_H \right)^2 \left( 500 \mbox{GeV} \over m_{\tilde a}\right),
 \label{YaDM}
\end{eqnarray}
for $g_*=100$ and $z_c= m_{\tilde a}/T_c=5$. This is more than $10^4$ times larger than the required value for a right dark matter density. If some decay mode is kinematically allowed before the electroweak symmetry breaking, such as $\tilde a \to H \tilde H$, the integration range of $z=[0, \infty]$ in Eq.~(\ref{YahH}) can be taken to increase the axino population by factor of 6.
 Thus, we can safely conclude that the Higgsino dark matter population from the DFSZ axino decay is large enough to require re-annihilation.

Next question is whether the decay temperature of the axino is
larger than the free-out temperature of the dark matter. The axino
decay process $\tilde a \to h \tilde H $ ($\tilde H$ may decay
further to the LSP if it is not the LSP.) has the decay rate:
\begin{equation}
 \Gamma_{\tilde a} \approx {c_H^2\over 16\pi} \left( \mu \over f_a \right)^2 m_{\tilde a},
\end{equation}
which corresponds to the decay temperature $T_D = g_*^{-1/4} \sqrt{ 3 \Gamma_{\tilde a} M_P}$.  Defining $x_D\equiv m_\chi/T_D$ where $\chi$ denotes the LSP (the lighter Higgsino in our case),
one gets
\begin{equation} \label{xD}
 x_D \approx 34\left( g_* \over 70\right)^{1/4}
 \left( 500 \mbox{GeV} \over m_{\tilde a}\right)^{1/2}
 \left( m_{\chi} \over \mu \right) \left( f_a /c_H\over 10^{11}\mbox{GeV}\right).
\end{equation}
Thus, requiring the decay temperature smaller than the freeze-out temperature $T_f$, that is,
$x_D > x_f$ where $x_f\equiv m_\chi/T_f\approx 25$, one needs $f_a \gtrsim 7\times10^{10}$GeV.
Having $T_D < T_f$, the Boltzmann equation for the LSP abundance can be solved as \cite{Choi08}
\begin{equation} \label{Y-reann}
 Y_{\chi}^{-1} \approx  Y_{\chi}^{-1}(T_D) + {\langle \sigma_A v\rangle s(T_D) \over H(T_D)},
\end{equation}
where the first term $Y_{\chi}(T_D)^{-1} \approx Y_{\tilde a}^{-1} $, as given in Eq.~(\ref{YaDM}),
is much smaller than the second term.   Thus, the dark matter relic density is determined by
\begin{equation} \label{Omega-reann}
 \Omega_{\chi} h^2 = {10^{-11}\mbox{GeV}^{-2} \over (g_*/70)^{1/2} }
 {x_D \over \langle \sigma_A v\rangle },
\end{equation}
which requires the dark matter annihilation rate: $\langle \sigma_A v\rangle =x_D\times 10^{-10} \mbox{GeV}^{-2}$.
From the relation (\ref{xD}), we get $x_D\approx 340$ for the (approximate) upper end of
$f_a/c_H=10^{12}$GeV and the lower value of $x_D$ is put by $x_f\approx 25$. Therefore, we conclude that the annihilation rate in the range of
\begin{equation} \label{ann-rate}
 2.5\times10^{-9}\mbox{GeV}^{-2} \lsim \langle \sigma_A v\rangle
 \lsim 3.4\times 10^{-8} \mbox{GeV}^{-2}
\end{equation}
is required for the neutralino LSP.  Such a large annihilation
rate can be obtained indeed for the Higgsino-like LSP as can be seen
from the general parameter scan made in
Ref.~\cite{Edsjo97} including co-annihilations.
In our case, co-annihilation does not occur as
the Higgsinos are not in thermal equilibrium and thus the heavier
states decay away to the lighter ones.


\medskip

For the calculation of the neutralino LSP annihilation rate
in the mass range $200\sim500$ GeV, we
take the leading s-wave contributions in the decoupling limit of
heavy Higgs bosons and all the sfermions. All the relevant terms
can be found in Ref.~\cite{Jungman95} and let us recollect them in
the below. We use the following convention for the diagonalization
of the chargino and neutralino mass matrices:
\begin{eqnarray}
 \chi^-_n &=& U_{n1} \tilde{W}^- + U_{n2} \tilde{H}_1^-, \\
 \chi^+_n &=& V_{n1} \tilde{W}^+ + V_{n2} \tilde{H}_2^+, \nonumber\\
 \chi^0_n &=& N_{11} \tilde B + N_{12} \tilde W^3 + N_{13} \tilde H^0_1 + N_{14} \tilde H^0_2.
 \nonumber
\end{eqnarray}
Note that we use the notation $\chi\equiv \chi_1^0$.  The s-wave contributions to the annihilation rate in the $W^\pm W^\mp$, $Z^0 Z^0$, $Z^0 h$ and $t\bar{t}$ final states are given by
\begin{eqnarray} \label{WW}
\langle \sigma_A v\rangle_{WW} &=&
  \pi \alpha_2^2 {\beta_{WW}^{3/2}\over m_\chi^2}
  \left[ \sum_{n=1}^2
  {\mathbf{U}_n^2 + \mathbf{V}_n^2 \over 1 + r^2_{\chi^\pm_n} - r_W^2}
  \right]^2, \\
\langle \sigma_A v\rangle_{ZZ} &=&
  {\pi \alpha_2^2 \over 8 c_W^4}
  {\beta_{ZZ}^{3/2}\over m_\chi^2}
  \left[ \sum_{n=1}^4
  {\mathbf{Z}_n^2 \over 1 + r^2_{\chi^0_n} - r_Z^2}
  \right]^2, \nonumber\\
\langle \sigma_A v\rangle_{Zh} &=&
  {\pi \alpha_2^2 \over  c_W^2}
  {\beta_{Zh}^{3/2}\over m_\chi^2}
  \left[ \sum_{n=1}^4
  {-\mathbf{Z}_n \mathbf{T}_n (m_{\chi^0_n}-m_\chi) \over
  m_Z(1 + r^2_{\chi^0_n} - r_Z^2/2 - r_h^2/2)}
  +{\mathbf{Z}_0 m_\chi^2 \over c_W m_Z^2}
  \right]^2, \nonumber\\
\langle \sigma_A v\rangle_{tt} &=&
  {3\pi \alpha_2^2 \over 16 c_W^4}
  {\beta_{tt}^{1/2}\over m_\chi^2}
  \left[
  \mathbf{Z}_0 {m_t m_\chi \over m_Z^2}
  \right]^2,  \nonumber
\end{eqnarray}
where $r_{X} \equiv m_{X}/m_\chi$ and $\beta_{XY}\equiv\sqrt{1-(r_X^2+r_Y^2)/2-(r_X^2-r_Y^2)/4}$.
The couplings appearing in the above equations are defined as follows:
\begin{eqnarray}
 \mathbf{U}_n &=& -{1\over\sqrt{2}} N_{13} U_{n2} + N_{12} U_{n1}, \\
 \mathbf{V}_n &=& -{1\over\sqrt{2}} N_{14} V_{n2} + N_{12} V_{n1}, \nonumber\\
 \mathbf{Z}_n &=& -N_{13}^2 + N_{14}^2 \nonumber\\
 \mathbf{T}_n &=& {1\over2} (c_\beta N_{13} - s_\beta N_{14})(N_{n2}-t_W N_{n1})
          + (1\leftrightarrow n). \nonumber
\end{eqnarray}

The LSP annihilation rate is shown in Fig.~1 by blue solid contours in the $\mu$--$M_1$ plane assuming $M_2=2 M_1$ and $t_\beta=10$.  Remark that the annihilation rates for the whole Higgsino-like LSP region (and the bino-like region with a large mixture of Higgsino) are in the range of $(1\sim3)\times10^{-8}$GeV$^{-2}$ which fits well into the desired range for re-annihilation (\ref{ann-rate}).  As can be expected, the most bino-like LSP region with a smaller Higgsino mixture ($\mu \gg M_1$) is completely excluded as the annihilation rate becomes
smaller than about $2\times 10^{-9}$ GeV$^{-2}$ making  $\Omega_{\chi}h^2> 0.11$.
In the same figure also shown is the spin-independent nucleonic cross-section which is calculated using the following formula \cite{Jungman95}:
\begin{equation}
 \sigma_{SI} = 4\pi \alpha_2^2 {\mu_n^2 m_n^2 \over m_W^2 m_h^4} \mathbf{T}_1^2 f_{T}^2
\end{equation}
where $\mu_n$ is the reduced mass, $m_n$ is the nucleon mass, and $f_T$ is the nuclear form factor
$f_{T} \equiv  f_{T_{lq}} + 2 f_{T_G}/7$ with $f_{lq}=\sum_{q=u,d,s} f_{T_q}$ and $f_{T_G}=1-f_{T_{lq}}$.
Note that the direct detection rate has a large uncertainty due to the
nucleon form factor uncertainty (see, e.g., \cite{Bottino99}). In our numerical calculation, we choose $f_T=0.27$ adopting the values of $f_{T_u}+f_{T_d}=0.05$, and  $f_{T_s}=0.013$ from the recent lattice calculation \cite{JLQCD10}.

The nucleon--LSP scattering depends sensitively on the bino fraction in $\mathbf{T}_1$ and the lightest Higgs boson mass.
For the Higgs mass $m_h=115$ GeV, the nucleonic scattering cross-section is shown to be
as large as about $2\times10^{-44}$cm$^2$ for the Higgsino-like LSP with a large bino mixture
in the mass range $m_\chi= (170\sim190)$ GeV.

\begin{figure}
\begin{center}
\includegraphics[width=0.55\linewidth]{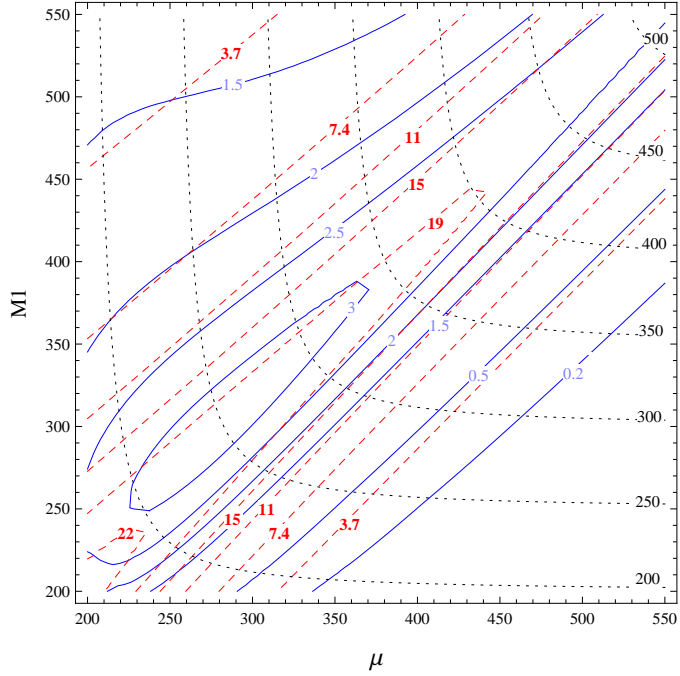}
\end{center}
\caption{In the $\mu$--$M_1$ plane showed is the LSP annihilation
rate (blue solid contours) and the spin-independent
nucleonic scattering cross-section (red dashed contours)
in the unit of $10^{-8}$GeV$^{-2}$ and $10^{-45}$cm$^2$,
respectively.  The dotted lines show the LSP
mass. The calculation is made with the lightest Higgs mass 115 GeV,
assuming $M_2=2 M_1$ and the decoupling limit
of heavy Higgs bosons and all the sfermions.
The nuclear form factor is taken to be  $f_{T}=0.27$.
}\label{fig}
\end{figure}

\medskip


Before closing this section, let us make remarks on the impact of the saxion $s$,
the scalar partner of the axion \cite{Kim91}, whose mass is expected to be order
of the supersymmetry breaking scale $\sim 10^{2-3}$ GeV.
In the DFSZ axion model \cite{Baer10}, the dominant interaction terms of the saxion contain
\begin{equation} \label{Ls}
{\cal L}_s = {s\over f_a} \left[
{\sqrt{2} x \over 2 }   \partial_\mu a \partial^\mu a
+ c_H \mu (\tilde H_1 \tilde H_2 + h.c.)
+ { c'_H \mu^2 } {h^2 \over2} \right]
\end{equation}
where the Heavy Higgs bosons are decoupled. Here, $x\equiv \sum_i q_i^3 v_i^2/f_a^2$
depends on the charges $q_i$ and the vacuum expectation values $v_i$
of the Peccei-Quinn symmetry breaking fields $S_i$ \cite{Chun95}, and $c'_H$ parameterizes
the dependence on soft supersymmetry breaking terms. Note that the axion scale is given by
$f_a = \sqrt{\sum_i q_i^2 v_i^2}$. Like the axino, the saxion population can be produced thermally from the second and third terms in Eq.~(\ref{Ls}).
When the decay of the saxion to the Higgsinos, $s \to \tilde H \tilde H$, is allowed,
its inverse decay determines the thermal saxion population as
\begin{eqnarray} \label{YsHH}
 Y_{s} &\approx& {405  \over 8 \pi^3 g_*^{3/2}} {\Gamma_{s\to \tilde H \tilde H} M_P \over m_{s}^2}  \nonumber\\
 &\approx& 2.5\times 10^{-6}\left(\mu \over 200 \mbox{GeV}\right)^2
 \left( 10^{11} \mbox{GeV} \over f_a/c_H \right)^2 \left( 500 \mbox{GeV} \over m_{s}\right),
 \label{YsHH}
\end{eqnarray}
with $g_*=100$ and the decay rate $\Gamma_{s\to \tilde H\tilde H}=c_H^2 (\mu/f_a)^2 m_s/4\pi$.
If the Higgsino channel is not open, the saxion decay to the Higgs bosons,
$s \to hh$, from the third term in Eq.~(\ref{Ls}) becomes the main source
for the thermal saxion production which leads to the saxion abundance
comparable to Eq.~(\ref{YaDM}).
The saxion population can arise also from a coherent oscillation driven
by an initial misalignment, $s_i$:
\begin{equation}
 Y_s \approx 6.7\times10^{-7} \left( 500 \mbox{GeV}\over m_s\right)^{1/2}
 \left( s_i \over 10^{12} \mbox{GeV}\right)^2
 \label{Ysi}
\end{equation}
if the reheat temperature is high enough \cite{Chang96}.  When the reheat temperature is low, 
$T_R \lsim 10^9$ GeV,
the coherent production is typically  smaller than the thermal production (\ref{YsHH}).
The contribution of Eq.~(\ref{Ysi}) can be larger than the thermal contribution (\ref{YsHH})
if $s_i \sim f_a \sim 10^{12}$ GeV, and thus the LSP dark matter
can mostly come from the saxion decay.

Contrary to the axino case, however, the overproduction problem of the LSP dark matter from the saxion decay is model-dependent. First of all, there is no such a problem when
the saxion decay to the LSP is forbidden kinematically.  Although it is allowed, the saxion may mainly decay to the axions through the first term in Eq.~(\ref{Ls}) giving the decay
rate $\Gamma_{s\to aa}= x^2 m_s^3/64\pi f_a^2$.  In this case, the produced axions are red-shifted away and the LSP dark matter production is suppressed by the factor
$\Gamma_{s\to \tilde H \tilde H}/\Gamma_{s\to aa} \approx 16 c_H^2 \mu^2/x^2 m_s^2$.

\section{Conclusion}

Dark matter property is discussed in the Kim-Nilles mechansim which realizes the DFSZ axion solution to the strong CP problem and solves the $\mu$ problem at the same time.
The DFSZ axino regeneration by the $\mu$ term interactions (which is independent of the reheat temperature) is more efficient than the KSVZ axino regeneration from the QCD interaction at the reheat temperature of order TeV, and thus puts also a stringent bound on the stable axino mass. The DFSZ axino dark matter mass can vary from ${\cal O}$(0.1 keV) and ${\cal O}$(10 MeV) depending on the allowed decay processes and the axion scale $f_a\approx 10^{10}\sim 10^{12}$ GeV. When the stable axino has the mass of order the electroweak scale, the reheat temperature has to be low enough to suppress the regeneration rate and, of course, dark matter can consist of heavy axinos
if the reheat temperature is appropriately adjusted.

The DFSZ axino may decay to a neutralino through the $\mu$ term and the overproduced neutralinos can be thermalized or re-annihilate sufficiently to saturate the observed dark matter density. A light Higgsino-like LSP  is shown to have the annihilation rate $(1\sim3)\times10^{-8}$GeV$^{-2}$ and thus becomes a good dark matter candidate realizing the re-annihilation scenario  for the axion scale
$f_a\approx 10^{11\sim 12}$ GeV.  Furthermore, the Higgsino-like LSP with a sizable mixture of the bino component has the spin-independent nucleonic cross-section as large as $2 \times 10^{-44}$cm$^2$ for the Higgs mass 115 GeV, and thus might be observed in the future direct detection experiments.

\medskip

{\bf Acknowledgments:} This work was supported by Korea Neutrino
Research Center through National Research Foundation of Korea
Grant (2009-0083526).  The author thanks Ki-Young Choi for discussions 
and careful reading of the manuscript.

\end{document}